\documentclass{article}
\usepackage{spconf,amsmath,graphicx, amsfonts}
\usepackage[sort,nocompress]{cite}

\usepackage{booktabs}
\title{Robust Disentangled Variational Speech Representation Learning for Zero-shot Voice Conversion}
\name{Jiachen Lian$^{1,2}$\thanks{Work done when Jiachen was an intern at Tencent AI Lab, Bellevue, WA}, Chunlei Zhang$^{2}$,  Dong Yu$^{2}$}
  
  \address{$^{1}$ UC Berkeley, EECS, CA $^{2}$ Tencent AI Lab, Bellevue, WA\\
  \small \tt jiachenlian@berkeley.edu, \{cleizhang, dyu\}@tencent.com
  }
\begin{document}
\ninept
\maketitle

\begin{abstract}
Traditional studies on voice conversion (VC) have made progress with parallel training data and known speakers. Good voice conversion quality is obtained by exploring better alignment modules or expressive mapping functions. In this study, we investigate zero-shot VC from a novel perspective of self-supervised disentangled speech representation learning. Specifically, we achieve the disentanglement by balancing the information flow between global speaker representation and time-varying content representation in a sequential variational autoencoder (VAE). A zero-shot voice conversion is performed by feeding an arbitrary speaker embedding and content embeddings to the VAE decoder. Besides that, an on-the-fly data augmentation training strategy is applied to make the learned representation noise invariant. On TIMIT and VCTK datasets, we achieve state-of-the-art performance on both objective evaluation, i.e., speaker verification (SV) on speaker embedding and content embedding, and subjective evaluation, i.e., voice naturalness and similarity, and remains to be robust even with noisy source/target utterances. 
\end{abstract}
\begin{keywords}
Self-supervised Disentangled representation learning, zero-shot style transfer, voice conversion, variational autoencoder 
\end{keywords}
\vspace{-1ex}
\section{Introduction}
\label{introduction}

Voice Conversion (VC) seeks to automatically convert the non-linguistic information of a source speaker to a target speaker, while keeping the linguistic content unchanged. The non-linguistic information may include timbre (i.e., speaker identity), emotion, accent or rhythm, to name a few. Due to its potential for applications in privacy protection, security and entertainment industry etc.~\cite{sisman2020overview,bahmaninezhad2018convolutional,zhang2020durian}, VC has received long-term research interest.

%Based on the conditions of source and target speakers that a VC system can access in the training phase, we can categorise current VC approaches into one-to-one, many-to-one, many-to-many, any-to-many and any-to-any (i.e., zero-shot) VC. Conventional studies focus on one-to-one VC, which requires parallel data between a pair of source-target speakers. In these systems, acoustic features of source and target utterances are firstly aligned frame-wise with an alignment module, and a conversion transformation is trained to map time-aligned source acoustic features to target features. The requirement of parallel data limits the application of such models in the real world. Recent

We can categorise current VC systems into two methodologies. The first one employs a conversion model to map source acoustic features to target acoustic features~\cite{sisman2020overview,stylianou1998continuous,toda2007voice} . For the conventional VC approaches with parallel training data, acoustic features are first extracted from the source and target utterances. Then, the acoustic features are aligned frame-wise with an alignment module~\cite{berndt1994using}. Studies have shown that the alignment step can be bypassed through using sequence-to-sequence models for direct source-target acoustic mapping with better VC performance~\cite{zhang2019sequence}. For direct mapping VC with nonparallel training data, progress has been made with generative adversarial networks (GAN) based many-to-many VC systems~\cite{kameoka2018stargan,kaneko2018cyclegan}. Although widely investigated, the direct mapping method assumes that the speaker of source-target VC pair is pre-known, which limits the application of such models in the real world. To relax this constraint, the second methodology constructs VC based on explicitly learned speaking style and content representations. Among these approaches, phonetic posteriorgrams (PPGs) is widely used as the speaker independent content representations~\cite{sun2016phonetic,guo2020phonetic}, and speaker embeddings extracted from a pre-trained speaker verification model are often assumed to carry timbre information~\cite{zhang2021transfer}. They have been successfully applied to tasks such as many-to-many VC or any-to-many VC. For zero-shot VC, both AUTOVC and AdaIN-VC construct encoder-decoder frameworks~\cite{autovc,adaIN}. The encoder compress the speaking style and the content information into the latent embedding, and the decoder generate a voice sample by combining a speaking style embedding and a content embedding. To achieve a better VC performance, these models require positive pair of utterances (i.e., two utterances come from the same speaker) during training, and the systems have to rely on pre-trained speaker models. 

In this study, we focus on the problem of speaker identity conversion. We extend VAE as the backbone framework for learning disentangled content representation and speaking style representation, where balanced content and style information flow is achieved in the VAE training~\cite{vae}. We show that the vanilla VAE~\cite{vae, dsvae} loss can be extended to force strong disentanglement between speaker and content components, which is intuitively explained from three levels. 
% Inspired by AdaIN-VC, where a 1D instance normalization is applied across time to better remove the speaker related information \cite{adaIN}. We find that applying 2D instance normalization to the early sharing layer of VAE encoder is better to balance the content and speaking style flow. 
In addition, we explore to make the learned representation robust against background noise/music and interfering speaker etc. An on-the-fly data augmentation is introduced as the inductive bias to the VAE training, with this training strategy, we arrive at a denoising disentangled sequential VAE (D-DSVAE), where low quality speech input is allowed to test for VC. With all these contributions, our proposed system achieves the state-of-the-art VC performance and improved robustness.

\vspace{-3ex}
\section{Proposed Methods} \label{proposed method}
\begin{figure}[tbp]
    \centering
    \includegraphics[width=0.6\linewidth, height=0.45\linewidth]{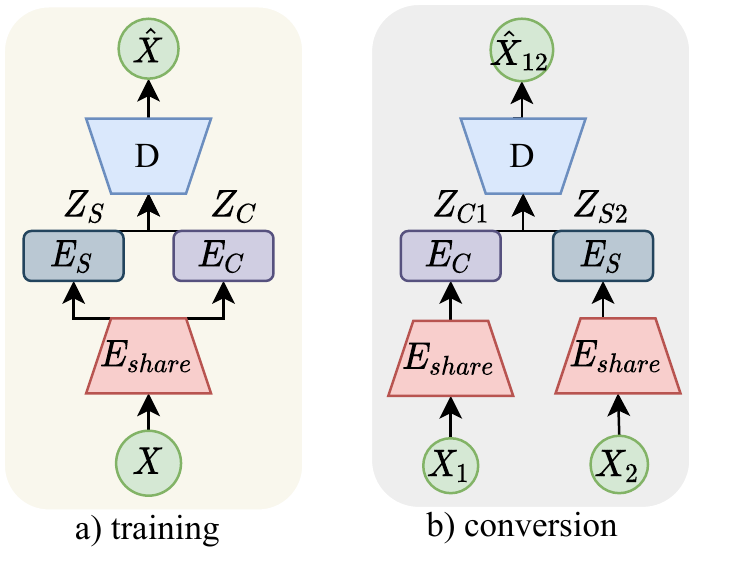}
        \vspace{-3ex}
    \caption{A flow diagram of proposed VC system.}
    \label{fig:se}
\end{figure}

\vspace{-1ex}
We start with introducing some notations. Denote a speech segment variable $X=[x_{1}, x_{2},...,x_{T}]$, which is the STFT or Mel-Spectrogram in our implementation. Denote $Z_S\in \mathbb{R}^{d_S}$ as speaker style latent representation and $Z_C=[z_{c1}, z_{c2}, ..., z_{c_T}]\in \mathbb{R}^{d_{C}\times T}$ as speech content latent representations. Denote $\theta$ as model parameters. The proposed VC system adopts the modified form of DSVAE \cite{dsvae} as our backbone, which is shown in Fig.\ref{fig:se}. The model takes $X$ as input, which is first passed into a shared encoder $E_{share}$ and it gives $W=[w_1,w_2,...,w_T]\in \mathbb{R}^{d_W\times T}$. Here $d_S$, $d_C$ and $d_W$ are positive integers. Then a speaker encoder $E_{speaker}$ and a content encoder $E_{content}$ take $W$ as input and model the posterior distribution $q_{\theta}(Z_S|W)=q_{\theta}(Z_S|X)$ and $q_{\theta}(Z_C|W)=q_{\theta}(Z_C|X)$ respectively. $Z_S$ and $Z_C$ are then obtained via sampling from $q(Z_S|X)$ and $q(Z_C|X)$. The expectation is that $Z_S$ only encodes speaker information and $Z_C$ only encodes content information. During the generation stage, the concatenation of $Z_S$ and $Z_C$ are passed into a shared decoder $D_{share}$ to generate the spectrogram $\hat{X}$, i.e. $\hat{X}=D(Z_S, Z_C)$. A vocoder is then applied to convert $\hat{X}$ to waveform. When performing voice conversion from source $X_1$ to target 
$X_2$, and the converted speech is $\hat{X}_{12}=D(Z_{S2}, Z_{C1})$, where $Z_{S2}$ and $Z_{C1}$ are sampled from $q(Z_S|X_2)$ and $q(Z_C|X_1)$ respectively, as shown in Fig.\ref{fig:se}. In the following sections, we first highlight the probabilistic graphic models in Sec.\ref{probabilistic graphic models}. Then, the objective function is introduced in Sec.\ref{loss objectives}, then we discuss its validity via proposing three sufficient conditions for achieving disentanglement between $Z_S$ and $Z_C$.

% \begin{figure}[h] \label{train_model}
% \begin{center}
% %\framebox[4.0in]{$\;$}
% \hbox{\hspace{-1.2em}\includegraphics[width=15cm]{iclr2022/train_model.png}}
% \end{center}
% \caption{Model}
% \end{figure}

\label{sec:model}

\subsection{Disentanglement-Aware Probabilistic Graphical Models} \label{probabilistic graphic models}
The frequently used method to achieve disentanglement between two components is to let these two components be probabilistically independent with each other. Following such intuition, we factorize both the prior distributions and posterior distributions of $Z_S$ and $Z_C$ by following independence assumption, which is consistent with \cite{dsvae, s3vae}. We denote $Z=[Z_S, Z_C]$ as a joint latent representation.
\vspace{-2ex}
\paragraph*{Prior}  The joint prior distribution is factorised as follows:
\begin{equation} \label{prior independence} 
    p_{\theta}(Z)=p(Z_S)p_{\theta}(Z_C)=p(Z_S)\prod_{t=1}^{T}P_{\theta}(z_{ct}|z_{<t})
\end{equation}
where $p(Z_S)$ is a standard normal distribution and $p_{\theta}(Z_C)$ is modeled by an autoregressive LSTM.
\vspace{-2ex}
\paragraph*{Posterior}
The joint posterior distribution is factorised as follows:
\begin{equation} \label{posterior independence}
\begin{split}
         q_{\theta}(Z|X)&=q_{\theta}(Z_S,Z_C|W)=q_{\theta}(Z_S|W)q_{\theta}(Z_C|W)
        % &=q_{\theta}(Z_S|W)\prod_{t=1}^T q_{\theta}(z_{ct}|w_{<t})
\end{split}
\end{equation}
Here, $q_{\theta}(Z_S|W)$ and $q_{\theta}(Z_C|W)$ are modeled by two independent LSTMs.  Implementation Detailes can be found in \ref{implementation details}.

\subsection{Loss Objectives} \label{loss objectives}
We first provide the loss objective as shown in Eq. \ref{vaeloss0}, which is simply the vanilla VAE \cite{vae} loss. After that, we will explain how it is related to disentanglement and how the disentanglement is achieved. 
\begin{equation}\label{vaeloss0}
\begin{split}
    &\mathcal{L}=\mathbb{E}_{p(X)}\mathbb{E}_{q_{\theta}(X|Z_{S}, Z_{C})}[-log(p_{\theta}(X|Z_S,Z_C))]+\\
    &\mathbb{E}_{p(X)}[\alpha kl(p(Z_S)||q_{\theta}(Z_S|X))+\beta kl(p_{\theta}(Z_C)||q_{\theta}(Z_C|X))]
\end{split}
\end{equation}
where $kl(.,.)$ denotes KL divergence between two distributions, $\alpha$ and $\beta$ are two balancing factors. The loss objective does not directly enforce the disentanglement between $Z_S$ and $Z_C$, however, we will intuitively explain in the following sub-sections how the disentanglement could be achieved from three levels.

\subsubsection{Variational Mutual Information and KL Vanishing in VAE} \label{lower bounded vae}
 In the original form of VAE \cite{vae}, the mutual information $MI(X,Z)=kl(p(X,Z)||p(X)p(Z))$ has its variational form $\hat{MI}_{\theta}(X,Z)=kl(q_{\theta}(X,Z)||p(X)p(Z))=\mathbb{E}_{p(X)}[kl(q_{\theta}(Z|X)||p(Z))]$, which we call the variational mutual information. The objective of VAE can be reformulated as:
\begin{equation}\label{vaeloss1}
    \mathcal{L}_{VAE}(\theta)=\mathbb{E}_{p(X)}\mathbb{E}_{q_{\theta}(Z|X)}[-log(p_{\theta}(X|Z))]+\hat{MI}_{\theta}(X,Z)
\end{equation}
The first term in RHS of Eq. \ref{vaeloss1} is the reconstruction loss. It is observable that minimizing both two terms are typically not achievable considering that lower reconstruction loss means higher variational mutual information between $X$ and $Z$, by assuming that we already achieve a good variational estimator, i.e. $q_{\theta}(Z|X)\approx p(Z|X)$. While there are exceptions when the decoder is designed to have powerful self-supervised generation capacity \cite{klvanishing}, these cases are not taken into consideration since the decoder is not such powerful in our scenario. Thus, VAE aims to reach a balance between the reconstruction loss and variational mutual information. 
It follows that the variational mutual information is lower bounded:  
\begin{equation} \label{lower bounded}
    \inf_{\theta^*} \hat{MI}_{\theta^*}(X,Z) \geq A(X,Z)>0
\end{equation}
where $\theta^*$ denotes a local optimal solution, $A(X,Z)$ is denotes the bounded variational mutual information, which is similar to $I_C$ in Eq. 2 in \cite{information-bottleneck}. The above formula guarantees that KL vanishing \cite{klvanishing} will not happen during VAE training, which is the foundation of disentanglement.

\subsubsection{Information Flow Between Multiple Latent Variables in VAE} \label{information flow section}
We argue that balanced information flow between $Z_S$ and $Z_C$ leads to disentanglement. Based on the Eq. \ref{vaeloss1}, the loss objective proposed in Eq. \ref{vaeloss0} can be formulated as:
\begin{equation}\label{vaeloss2}
\begin{split}
    \mathcal{L}(\theta)&=\mathbb{E}_{p(X)}\mathbb{E}_{q_{\theta}(X|Z_{S}, Z_{C})}[-log(p_{\theta}(X|Z_S,Z_C))]\\
    &+\alpha \hat{MI}_{\theta}(X,Z_S)+\beta \hat{MI}_{\theta}(X,Z_C)
\end{split}
\end{equation}
where the factorised form of \cite{dsvae} and \cite{s3vae} is a special case when $\alpha=\beta=1$. 
% Denote $\lambda=[\lambda_1,...,\lambda_k]$ as all $k$ hyperparameters(e.g. learning rate) in this work. Without loss of generality, we assume that $ \forall (\alpha, \beta, \lambda) \in \mathbb{H}=\{(\alpha, \beta, \lambda) | 0<\alpha\leq A_{\alpha}, 0<\beta \leq A_{\beta}, |\lambda_i|\leq A_{\lambda_i},i=1,...,k\}$, the model parameters $\theta$ will converge to its local minimal solution $\theta^*$, where $A_{\alpha}, A_{\beta}, A_{\lambda_i}$ are positive real numbers and they are guaranteed to exist. 
Following Sec. \ref{lower bounded vae}, the summation of information encoded by $Z_S$ and information encoded by $Z_C$ is sufficient enough to reconstruct the input speech, that is:
\begin{equation} \label{lower bounded for Z_S and Z_C}
\inf_{\theta^*} \{ \alpha \hat{MI}_{\theta^*}(X,Z_S)+\beta \hat{MI}_{\theta^*}(X,Z_C)\}\geq A(X,Z)>0
\end{equation}
Intuitively, by assuming convergence, when $\frac{\beta}{\alpha}$ is extremely large, the gradient of two variational mutual information terms will be dominated by $\hat{MI}(X,Z_C)$, which results in the vanishing of $KL(q_{\theta}(Z_C|X)||p(Z_C))$ so that the information encoded by the latent variable will be dominated by $Z_S$. Similar conclusion could be derived for the other side. The phenomenon of such imbalance is also observed in \cite{gradient-multi-task}. Theoretical expression of such gradient dominance is that, there exist $(\alpha_1, \beta_1), (\alpha_2, \beta_2)$ such that:
\begin{equation} \label{limit for infty}
 \alpha_1 \hat{MI}_{\theta^*}(X,Z_S)+\beta_1 \hat{MI}_{\theta^*}(X,Z_C)=\alpha_1 \hat{MI}_{\theta^*}(X,Z_S)
\end{equation}
\begin{equation} \label{limit for zero}
\alpha_2 \hat{MI}_{\theta^*}(X,Z_S)+\beta_2 \hat{MI}_{\theta^*}(X,Z_C)=\beta_2 \hat{MI}_{\theta^*}(X,Z_C)
\end{equation}
We experimentally observe that the variational mutual information loss is independent of reconstruction loss, i.e. $\alpha_1 \hat{MI}_{\theta^*}(X,Z_S) \approx \beta_2 \hat{MI}_{\theta^*}(X,Z_C)\approx A(X,Z)$.  In that sense, there is a double-sided information flow between $Z_S$ and $Z_C$. By choosing a proper $\frac{\beta^*}{\alpha*}$, it would be possible that there is no overlap between the information encoded by $Z_S$ and $Z_C$, i.e. $\hat{MI}_{\theta^*}(Z_S, Z_C)=0$. Eq. \ref{limit for infty} and Eq. \ref{limit for zero} will be experimentally justified in Sec. \ref{experiment}. 

\subsubsection{Time-invariant and Time-Variant Disentanglement} \label{average pooling}
We argue that applying average pooling techniques over the time dimension of a sequence of speech features leads to further disentanglement. According to Sec. \ref{information flow section}, even if $\hat{MI}(Z_S, Z_C)=0$, it is still undetermined which latent variable encodes speaker information or content information. Inspired by the 1D instancenorm method proposed in \cite{adaIN}, we attempt to keep the speaker information by only taking the average pooling on speech features to derive the posterior distribution $q(Z_S|X)$. By choosing $\alpha=\alpha*$ and $\beta=\beta*$ as indicated in the second sufficient condition, $Z_S$ will encode time-invariant information while 
$Z_C$ will encode time-variant information. Other normalizations techniques like instancenorm2D also help a lot to obtain better disentanglement performance, which will be discussed in Sec. \ref{experiment}.

\section{Experiments} \label{experiment}
\subsection{Dataset and Data Preprocessing}

\paragraph*{TIMIT}
We follow the official train/test split: 462 speakers for training and 24 speakers for testing \cite{timit}. For speaker verification (SV) task, we choose all possible trials from test set, which gives 18336 trials. 200 dimensional STFT features are extracted from a raw waveform with 25ms/10ms framing configuration. During training, the length of segment is fixed to 20 frames.
\paragraph*{VCTK}
For VCTK corpus \cite{vctk2017}, 90\% of the speakers are randomly selected for training and the remaining 10\% as for testing. For the evaluation of disentanglement performance in the SV task, we generate 36900 trials (22950 nontarget trials and 14040 target trials) from test set. We extract melspectrogram as features with a framing configuration of 64ms/16ms. The feature dimension is set to 80. We select a segment of 100 frames for the VAE training. 

\subsection{Model Architectures}
\paragraph*{Encoder, Decoder, Vocoder} 
There are two models designed for TIMIT and VCTK respectively. The encoder is composed of a shared encoder, speaker encoder and content encoder. (i) For TIMIT, the shared encoder is a 2-layer MLP with hidden size of 256. The content encoder is 2-layer BiLSTM with hidden size of 512, followed by a RNN layer with hidden size of 512. Then a 2-layer MLP of hidden size (512,64) is applied to model $q_{\theta}(Z_C|X)$. The speaker encoder is almost the same with content encoder except that there is an average pooling layer after RNN, and a 2-layer MLP is then applied to model $q_{\theta}(Z_S|X)$. The decoder is 1-layer MLP, followed by 2-layer BiLSTM which is followed by 2-layer MLP, where the hidden size is 256. Griffin-lim algorithm \cite{griffin1984signal} is applied as vocoder. (ii) For VCTK, modified from \cite{autovc}, the shared encoder is composed of three convolutional layers with 512 channels. Each convolutional layer is followed by a linear layer with dimension 512 and an Instancenorm2D layer \cite{instancenorm}. Also modified from \cite{autovc}, decoder includes a prenet with 512 channels and a postnet, which is a BiLSTM with hidden size of 512, followd by three convolutional layers with 512 channels, followed by a BiLSTM with hidden size of 512 and two separate linear layers to project the hidden dimension to 80 to model $p_{\theta}(X|Z_S,Z_C)$. A wavenet \cite{wavenet} pretrained on VCTK is used as Vocoder. Note that vocoder is only used for inference.

\paragraph*{Prior Distribution}
The prior distribution includes $p(Z_S)$ and $p_{\theta}(Z_C)$. $p(Z_S)$ is modeled by a Normal distribution $N(0, \mathbb{I}_{d_S})$, where there are no parameters. $p_{\theta}(Z_C)=\prod_{t=1}^{T}P_{\theta}(Z_{Ct}|Z_{C\tau<t})$ is modeled by an auto-regressive LSTM with hidden size of 512. Each hidden cell is followed by two one-layer MLPs with hidden size 512 to model $P_{\theta}(Z_{Ct}|Z_{C\tau<t})$, from which $Z_{Ct}$ is sampled and passed into the next LSTM cell.

% \begin{table}[h!] 
% \centering
% \begin{tabular}{|c| c c c c|} 
%  \hline
% $dim(\mu_S, \mu_C)$ & (256,256) & (128,128) & (64,64) & (32,32)  \\ [0.5ex] 
%  \hline
%  $EER\%$ on $\mu_S$  & 4.61 & 2.95 & 3.25 & 3.41\\ 
%  $EER\%$ on $\mu_C$ & 33.12 & 31.11 & 38.83 & 37.90\\ [1ex] 
%  \hline
% \end{tabular}
% \caption{EER on TIMIT test trials, fix $\beta/\alpha$ to be 10 and $\beta$ to be 1, and vary the dimension of $\mu_S$ and $\mu_C$}
% \label{timit embedding size eer}
% \end{table}

% \begin{table}[h!]
% \centering
% \begin{tabular}{|c| c c c c c|} 
%  \hline
% $\alpha$ & Col2 & Col2 & Col3 &  Col1 & Col2 \\ [0.5ex] 
%  \hline
%  $EER\%$ on $\mu_S$ & 6 & 87837 & 787 & 6 & 87837 \\ 
%  $EER\%$ on $\mu_C$ & 6 & 87837 & 787 & 6 & 87837 \\ [1ex] 
%  \hline
% \end{tabular}
% \caption{EER on VCTK test trials. Fix the latent dimension to be 64 and vary $\beta/\alpha$}
% \label{vctk alpha beta eer}
% \end{table}

% \begin{figure}[htb]
%     \centering % <-- added
% \begin{subfigure}{0.5\textwidth}
%  \hspace{-8pt}
%   \includegraphics[width=\linewidth]{ICASSP2022/unconditional/uncondition.png}
%   %\caption*{Speaker Identity and Gender Clustering}
%   \label{fi}
% \end{subfigure}\hfil % <-- added
% % \medskip
% % \begin{subfigure}{0.46\textwidth}
% %  \hspace{1pt}
% %   \includegraphics[width=\linewidth]{ICASSP2022/tsne/vctk_tsne2.png}
% %   \caption*{Speaker Gender}
% %   \label{fig:1}
% % \end{subfigure}\hfil % <-- added

% \caption{Unconditional Generation}
% \label{uncondition result}
% \end{figure}

\subsection{Implementation Details} \label{implementation details}
In the following, we first specify the hyper-parameters used for training, and then introduce evaluation metrics and training details. Next, the noise invariant method is introduced. Lastly, we detail the inference processes which include speaker verification, voice conversion.
\paragraph*{Hyper-parameters} Optimizer is adam with initial learning rate of 5e-4. Learning rate is decayed every 5 epochs with a factor of 0.95. Weight decay is 1e-4. Batchsize is 256. Both speaker dimension($d_S$) and content dimension($d_C$) are set to 64.  We performed grid search for $\alpha$ and $\beta$ and set $\alpha=1,\beta=20$ for TIMIT part and $\alpha=0.01, \beta=10$ for VCTK part.  The embedding size $d_S=d_C=64$.
\paragraph*{Evaluation Metrics}
(i) EER (Equal Error Rate) is used as the evaluation metrics for disentanglement. Lower EER on speaker embeddings and higher EER on content embeddings typically (would be clarified in the discussion section) indicate better disentanglement, as observed in \cite{fhvae, dsvae, s3vae}. (ii) We conduct a mean opinion score test to evaluate our proposed approach. Two different set are provided for MOS test. For both seen to seen VC and unseen to unseen VC, we select 6 speakers (3 females and 3 males), each speaker with one utterance. So 30 test cases are included in the set. The listener needs to give a score for each sample in a test case according to the criterion: 1 = Bad; 2 =
Poor; 3 = Fair; 4 = Good; 5 = Excellent. The final score for each model is calculated by averaging the collected results.

%\paragraph*{Training Details} The model is trained using Pytorch framework on NVIDIA Tesla P40. For both TIMIT and VCTK part, it takes around one GPU day for the loss to converge. We did not observe significant overfitting phenomenon during training and we just use the converged model for inference. 

\paragraph*{Noise Invariant training}
In the training of normal VAE, the VAE encoder takes clean acoustic feature $X$ as the input, and produce embedding $Z_S$ and $Z_C$. The decoder tries to reconstruct $X$ with the feeding information $Z_S$ and $Z_C$. To make the learned embedding $Z_S$ and $Z_C$ robust against background noise, we make a simple yet effective change in the training process. In the new data flow, a data augmentation module is applied to $X$, denoted as $X^{aug}$. The VAE encoder produces ${Z_S}^{aug}$ and ${Z_C}^{aug}$, which is the augmented version of $Z_S$ and $Z_C$. The decoding process $D({Z_S}^{aug}, {Z_C}^{aug})$ reconstructs acoustic feature $\hat{X}^{aug}$, and we maximize the likelihood of $\hat{X}^{aug}$ and clean reference $X$. By introduce this inductive bias, we expect the VAE framework not only disentangle global and local information, but also perform denoising at the latent space. In this study, clean utterance is augmented by MUSAN dataset \cite{snyder2015musan} with a balanced ``noise", ``music" and ``babble" distribution. 
\paragraph*{Inference Experiments}
Since the inference tasks are performed on the utterance-level, we also start with making some notations. Denote an utterance $U=[{X}^{(1)},{X}^{(2)},...,{X}^{(K)}]$, where ${X}^{(k)}$ is the kth segment. Denote ${Z_S}^{(k)}$ and ${Z_C}^{(k)}$ as the corresponding latent speaker and content variables for $X^{(k)}$. Here both ${Z_S}^{(k)}$ and ${Z_C}^{(k)}$ can be either the mean or a sample from $q({Z_S}^{(k)}|X^{(k)})$ and $q({Z_C}^{(k)}|X^{(k)})$ as we observed no significant difference made by these two. We implement two fundamental experiments: speaker verification and voice conversion. 

\begin{figure}[htb]
    \centering % <-- added
%\begin{subfigure}{\textwidth}
 \hspace{3pt}
  \includegraphics[width=\linewidth]{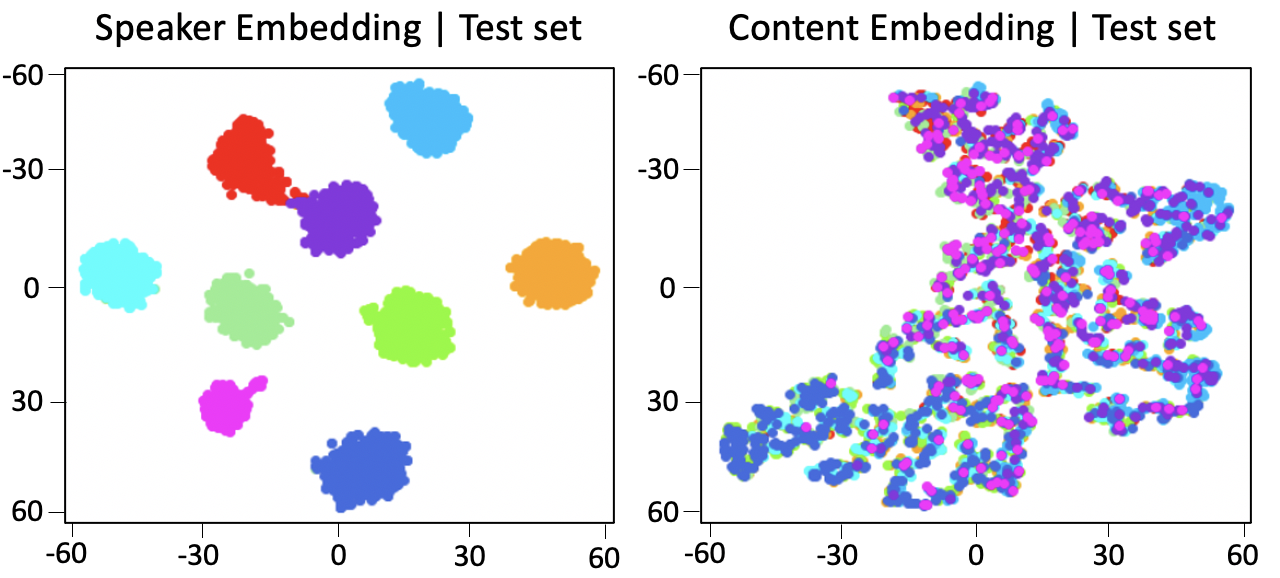}
  %\caption*{Speaker Identity and Gender Clustering}
  \label{fi}
%\end{subfigure}\hfil % <-- added
% \medskip
% \begin{subfigure}{0.46\textwidth}
%  \hspace{1pt}
%   \includegraphics[width=\linewidth]{ICASSP2022/tsne/vctk_tsne2.png}
%   \caption*{Speaker Gender}
%   \label{fig:1}
% \end{subfigure}\hfil % <-- added
\vspace{-5ex}
\caption{Visualization of Speaker and Gender Clustering on VCTK}
\label{vctk tsne}
\end{figure}

(i) Speaker Verification. We apply the same method with \cite{dsvae, s3vae} to derive the speaker and content embedding: $\mu_S(U)=(\Sigma_{k=1}^{K} {Z_S}^{(k)})/K$ and $\mu_C(U)=(\Sigma_{k=1}^{K}(\Sigma_{t=1}^{T}{z_c}_{t}^{(k)}/T))/K$. For TIMIT trials, we vary the value of $\frac{\beta}{\alpha}$ to observe the EER, and results are in Table \ref{timit alpha beta eer}.  We implement t-SNE \cite{tsne} visualization of speaker and content embeddings for VCTK, as shown in Fig. \ref{vctk tsne}. For speaker embedding, the cluster pattern is clear, while for content embedding, the evenly distributed scatters demonstrates that only limited (if not zero) speaker information is leaked in the content embedding.

(ii) Voice Conversion. Denote $U_{ij}$ as the converted utterance which takes $U_i$ as source and $U_j$ as target. We first obtain $\mu_S(U_j)$ as defined before. Second, pass 
${X_i}^{(k)}$ into the model and we will have ${Z_C}_i^{(k)}$. 
in generation stage, $D(\mu_{S}(U_j), {Z_C}_i^{(k)})$ gives $\hat{X_i}^{(k)}$; then $U_{ij}=[\hat{X_i}^{(1)},\hat{X_i}^{(2)},...,\hat{X_i}^{(K)}]$. We vary the value of $\frac{\beta}{\alpha}$ to observe the results of voice conversion. Since we do not observe big difference between VCTK and TIMIT parts, we just present the results for TIMIT in Fig. \ref{information flow} to illustrate the double-sided information flow as proposed in the second sufficient condition. 
% (iii) Unconditional Speech Generation. To testify how good $q(Z_S|X)$ is, instead of sampling the speaker embedding from $q(Z_S|X)$, we would sample it from $p(Z_S)$, which is $N(0, 0.5\mathbb{I}_{d_S})$ as mentioned before. However, according to the second sufficient condition, as long as we apply $\alpha^*$ and $\beta^*$(or we observe the good voice conversion performance), it is impossible that $q(Z_S|X)$ converges to $p(Z_S)$. Thus, we sample $Z_S$ from a specific set of $N(\mu_p, 0.5\mathbb{I}_{d_S})$. We randomly pick two utterances and only keep the $Z_C$. During the generation stage, we pick $\mu_p=\overrightarrow{-1},\overrightarrow{-0.5}, \overrightarrow{0}, \overrightarrow{0.5}$, $\overrightarrow{1}$ respectively and sample two $Z_S$ using the same random seed for each $\mu_d$ and we will obtain twenty utterances in total. $\overrightarrow{k}$ denotes a vector with entries of k. The results are shown in Fig. \ref{uncondition result}. 
\vspace{-2ex}
\vspace{-4ex}

\subsection{Results and discussions}
\begin{figure}[!ht]
    %\centering % <-- added
%\begin{subfigure}{\textwidth}
 \hspace{1pt}
  \includegraphics[width=\linewidth]{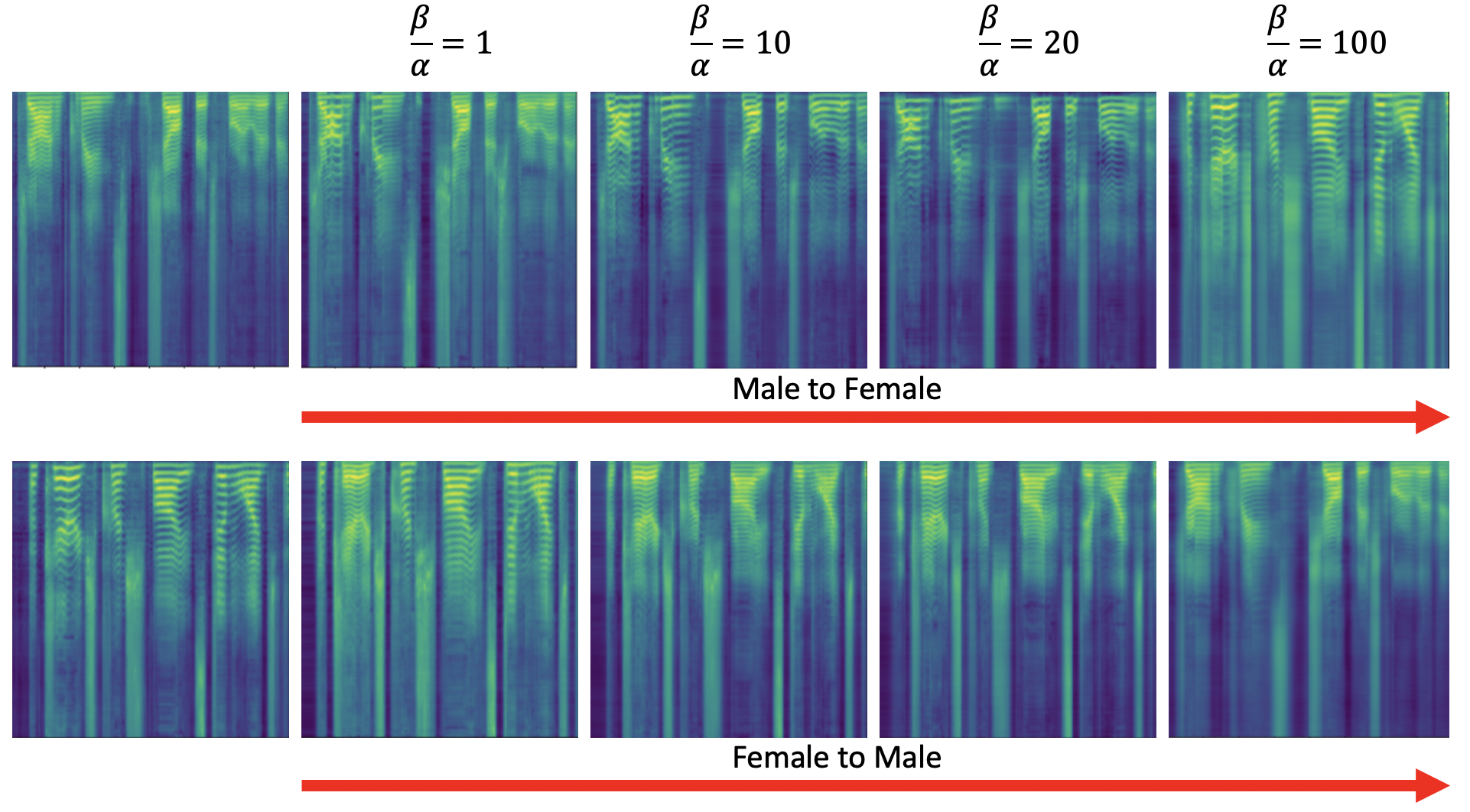}
  %\caption*{}
  \label{figggg}
%\end{subfigure}\hfil % <-- added
%\vspace{-15pt}
\caption{Visualization of Double-sided Information Flow. Four models are trained with four different $\frac{\beta}{\alpha}$. Given a pair of male and female voice, we perform double-sided voice conversion. For each row, the first figure is the reconstructed spectrogram. The remaining four figures are the converted spectrograms. For the purpose of presentation, we just use fixed length of segment for voice conversion}
\label{information flow}
\vspace{-2ex}
\end{figure}

\paragraph*{Disentanglement}
As observed in Fig.\ref{information flow}, only half of the utterance is converted to the target speaker for both two source utterances when $\frac{\beta}{\alpha}=1$, which indicates that $Z_S$ does not encode the whole speaker information and $Z_C$ encodes part of the speaker information. When $\frac{\beta}{\alpha}$ is increased to 10 or 20, the speaker EER becomes lower and content EER becomes higher, which indicates better disentanglement. Fig.\ref{information flow} also illustrates this point since $\frac{\beta}{\alpha}=10$ or $20$ successfully achieves voice conversion. When $\frac{\beta}{\alpha}=100$, Fig. \ref{information flow} indicates that instead of performing voice conversion, it actually does utterance swapping, which means that $Z_S$ encodes almost everything and $Z_C$ encodes almost nothing. When this case happens, it is still reasonable that speaker EER is low and content EER is high, as observed in Table \ref{timit alpha beta eer}. To this end, we can derive another conclusion that good eer is actually the necessary but not sufficient condition for good disentanglement for two reasons. First, $\frac{\beta}{\alpha}=1$ and $\frac{\beta}{\alpha}=100$ give similar EER, however, there is no disentanglement for the latter case. Second, $\frac{\beta}{\alpha}=10$ gives better EER, however, $\frac{\beta}{\alpha}=20$ gives better disentanglement if we take a closer look at middle part of the spectrogram in male to female conversion in Fig. \ref{information flow}. The better way to look at disentanglement is to look at both EER and information flow. Fig.\ref{information flow} shows that as $\frac{\beta}{\alpha}$ is increasing, $Z_S$ encodes more information and $Z_C$ encodes less information, which is consistent with the assumption in our proposed second sufficient condition. For VCTK trails, the best SV EER of $Z_S$ is 2.3\%, while the EER of $Z_S$ and $Z_C$ w.r.t. the best VC performance are 4.6\% and 44.5\%. The difference indicates that balancing the information flow for disentanglement in VC remains a challenging problem.  
%\vspace{-15pt}

\begin{table}[!ht]
\centering
\caption{EER (\%) for TIMIT test trials on varying $\frac{\beta}{\alpha}$.}.
\begin{tabular}{|c| c c c c c|} 
 \hline
$\frac{\beta}{\alpha}$ & 1 & 10 & 20 & 100 &DSVAE\cite{dsvae}\\ 
 \hline
  $\mu_S$  & 5.40 & 3.25 & 4.16 & 5.01 & 4.94 \\ 
  $\mu_C$ & 31.09 &38.83 & 37.16 & 38.79 & 17.49\\ 
 \hline
\end{tabular}

\label{timit alpha beta eer}
\end{table}
\vspace{-2ex}
\paragraph*{Voice conversion\footnote{Samples of voice conversion can be found at https://jlian2.github.io/Robust-Voice-Style-Transfer}} Table \ref{mos} shows the MOS results of different models. For fair comparison, we also generate VC samples from AUTOVC and AdaIN-VC, where we use their pretrained models with our generated VC pairs. As illustrated in the table, our proposed system outperforms AUTOVC and AdaIN-VC with a large margin in both naturalness and similarity. To test the effectiveness of noise invariant training, we conduct an additional MOS test. For test the noise invariant VC model, we apply background noise with signal to noise ratio range of 3-10 dB. As indicated in the MOS test, we only see marginal performance degradation compared with clean input, which shows the effectiveness of noise invariant training. 
\vspace{-1ex}
 \begin{table}[th]
     \scriptsize
          \caption{{ The results of the MOS (95\% CI) test on different
models.}}

     \centering
 
    \resizebox{8cm}{!}{
     \begin{tabular}{|c||c|c||c|c|}
     \hline
   & \multicolumn{2}{c ||}{\bf{seen to seen}} & \multicolumn{2}{c |}{\bf{unseen to unseen}}  \\
    model & naturalness & similarity  & naturalness & similarity\\
     \hline 
     \hline
     AUTOVC~\cite{autovc}  & 2.65$\pm$0.12  & 2.86$\pm$0.09 & 2.47$\pm$0.10 & 2.76$\pm$0.08    \\
     
     AdaIN-VC~\cite{adaIN} &2.98$\pm$0.09  & 3.06$\pm$0.07 & 2.72$\pm$0.11  & 2.96$\pm$0.09  \\
         
    Ours & 3.40$\pm$0.07  & 3.56$\pm$0.06 & 3.22$\pm$0.09  & 3.54$\pm$0.07   \\
    Ours(noisy) & 3.23$\pm$0.09  & 3.43$\pm$0.07 & 3.12$\pm$0.08  & 3.47$\pm$0.08   \\
     \hline 
     \end{tabular}
    }  
    \label{mos} 
    \end{table}
\vspace{-15pt}

% \paragraph*{Unconditional Generation}
% As shown in Fig. \ref{uncondition result}, unless the mean is the zero vector, the speaker identity is majorly determined by the mean of the prior distribution and the variance does not contribute too much on it. Note that sample 1 and sample 2 have almost the same identity. Furthermore, we observe that  positive $Z_S$ gives male speaker and negative $Z_S$ give female speaker. This natural boundary is also observed in Fig. \ref{vctk tsne}. This would partially explain the case of zero vector. When mean is zero, $Z_S=0+\epsilon*\sigma$ which randomly postive or negative and it follows that the speaker gender/identity is changed. Another point is that for both two utterances 

\section{Conclusion}
In this study, we proposed a novel zero-shot voice conversion system. By decomposing speech into speaker representation $Z_C$ and content representation $Z_S$ with a sequential VAE framework, the conversion can be as simple as: a) swapping the speaker representation; b) feeding source content embedding and target speaker embedding to the decoder to generate the acoustic features; c) convert the acoustic feature to the waveform with a vocoder. We improved the DSVAE framework by analyzing the information flow global between speaker representation and local content representation. Noise invariant training was investigated, which enabled the VC system to handle low quality speech input. For both SV and VC tasks, we achieved the state-of-the-art system performance on TIMIT and VCTK. We believe this study serves as a preliminary research work, and can be beneficial to other domains, such as speech recognition, text to speech studies.      

% Below is an example of how to insert images. Delete the ``\vspace'' line,
% uncomment the preceding line ``\centerline...'' and replace ``imageX.ps''
% with a suitable PostScript file name.
% -------------------------------------------------------------------------

%\vfill
%\pagebreak

\bibliographystyle{IEEEbib}
\bibliography{mybib}

\end{document}